\documentclass[10pt
]{article}
\usepackage[OE]{express}
\usepackage{graphicx}
\usepackage[T1]{fontenc}
\usepackage{amssymb}	
\usepackage{mathrsfs,amsmath}
\usepackage{units}
\usepackage{comment}
\usepackage{upgreek}

\begin{document}
\title{Towards fully commercial, UV-compatible fiber patch cords}

\newcommand{\nov}{ALPhANOV}
\newcommand{\nkt}{NKT Photonics}
\newcommand{\Ht}{H$_2$}
\newcommand{\Nt}{N$_2$}
\renewcommand{\d}[2][]{\frac{\text{d}^{#1}}{\text{d}#2^{#1}}}
\newcommand{\e}[1]{\text{e}^{#1}}

\author{Christian D. Marciniak,\authormark{1,*} Harrison B. Ball,\authormark{1} Alex T.-H. Hung,\authormark{1} and Michael J. Biercuk\authormark{1}}

\address{ARC Centre for Engineered Quantum Systems, School of Physics, The University of Sydney, NSW 2006, Australia}

\email{\authormark{*}christian.marciniak@sydney.edu.au}


\begin{abstract}
We present and analyze two pathways to produce commercial optical-fiber patch cords with stable long-term transmission in the ultraviolet (UV) at powers up to $\sim\unit[200]{mW}$, and typical bulk transmission between $\unit[66-75]{\%}$. Commercial fiber patch cords in the UV are of great interest across a wide variety of scientific applications ranging from biology to metrology, and the lack of availability has yet to be suitably addressed. We provide a guide to producing such solarization-resistant, hydrogen-passivated, polarization-maintaining, connectorized and jacketed optical fibers compatible with demanding scientific and industrial applications. Our presentation describes the fabrication and hydrogen loading procedure in detail and presents a high-pressure vessel design, calculations of required \Ht\ loading times, and information on patch cord handling and the mitigation of bending sensitivities. Transmission at $\unit[313]{nm}$ is measured over many months for cumulative energy on the fiber output of > $\unit[10]{kJ}$ with no demonstrable degradation due to UV solarization, in contrast to standard uncured fibers.  Polarization sensitivity and stability are characterized yielding polarization extinction ratios between $\unit[15]{dB}$ and $\unit[25]{dB}$ at $\unit[313]{nm}$, where we find patch cords become linearly polarizing.  We observe that particle deposition at the fiber facet induced by high-intensity UV exposure can (reversibly) deteriorate patch cord performance and describe a technique for nitrogen purging of fiber collimators which mitigates this phenomenon.
\end{abstract}







\section{Introduction}

Single-mode fiber patch cords are employed to meet the stringent demands of many physical science experiments - frequently even improving upon free-space performance by stabilizing against environmental fluctuations, protecting users against scattered radiation, and improving the spatial quality of the optical mode.  Polarization-maintaining (PM) fibers are also important in many atomic physics experiments and other applications where the polarization state of the input light must be preserved. 

There is a need for such fibers in many experimental settings, but also a fundamental incompatibility of the underlying technology with a growing set of applications requiring guided-mode operation in the ultraviolet.  In fields such as ion trapping\cite{Seidelin2006}, quantum computation\cite{Blatt2008}, metrology\cite{Schmoeger2015,Diddams2001,Ludlow2015}, spectroscopy\cite{Rosenband2007}, medicine\cite{Prahl1998}, and biology\cite{Bartels2015} the relevant optical transitions to be driven are often in the UV, and an expansion of source availability \cite{Lo2014,DiodeLasers2006,DiodeLasers2010,Carollo2017} has enhanced accessibility in these fields.  Nonetheless, there are no commercially available fiber patch cables compatible with power handling in the mW level below approximately $\unit[370]{nm}$.

The disparity between supply and demand arises due to radiation-induced photo darkening in silica optical fiber. This phenomenon, also known as UV solarization, causes optical fibers to gradually and permanently increase in opacity at the transmitted wavelength. The underlying physics is laser-induced formation of a range of colour centers\cite{Skuja2001} in the silica core. While solarization can occur in the visible and infrared as well, the threshold intensity required for noticeable degradation decreases with wavelength; UV solarization of silica fibers can occur at input powers as low as a few $\unit[]{\upmu W}$.  In such circumstances the optical intensity is relatively higher in the UV than the visible: the mode field diameter decreases proportional to wavelength and the overlap of the radiation with the absorption cross-section of relevant colour centers is much larger. Unfortunately there is an apparent zero-sum-game associated with simply increasing the mode field diameter, as this in turn sacrifices single-mode operation \cite{WaveguideTheory}. 

Recently, significant progress has been made on producing research-grade solarization-resistant UV fibers \cite{OH_Content, Colombe2014}, building on working knowledge of solarization immunization developed in the early 1990s \cite{Karlitschek1998,Karlitschek1998a}. Solarization immunity can be imparted by pressure loading silica fibers with molecular hydrogen and subsequently curing, {\emph i.e.} inducing a stable chemical change, through exposure to UV radiation. A complementary development in optical fiber technology came from the employment of large-mode-area (LMA) photonic crystal fibers (PCFs) which provide single-mode UV transmission with reduced intensity and facilitate efficient hydrogen loading in some configurations. These technical achievements must now be converted to commercially viable products capable of achieving demanding performance specifications with robustness appropriate for use outside of pristine laboratory conditions.  

In this paper, we describe work towards the realization of commercially manufactured single-mode, polarization-maintaining UV patch cords for scientific and industrial applications. We present two pathways to obtaining such patch cords which simplify production and exploit the reliable quality of established commercial fiber connectorization and jacketing expertise. We demonstrate stable, long-term transmission of UV light using fibers cured prior to connectorization and jacketing (pre-cured) and those cured as complete patch cables (post-cured), with peak power levels of up to $\unit[200]{mW}$ or cumulative radiant energy of > $\unit[10]{kJ}$, and typical bulk transmission between $\unit[66-75]{\%}$. We purge fiber collimators with dry \Nt\ in order to mitigate non-solarization (reversible) fiber degradation due to the accumulation of particles on fiber facets induced by the strong local field. Our tests characterize bend sensitivity of UV stabilized fiber patch cords and inform the development of a stretching routine to mitigate plastic memory which can destabilize transmission and even inhibit appropriate curing. Finally, we measure the polarization dependence of patch cord transmission, finding polarization extinction ratios ranging from $\unit[15-25]{dB}$ for post-cured patch cords, comparable to the best commercial PM fibers in the visible, and discovering that patch cords become linearly polarizing in the UV.

The remainder of this manuscript is organized as follows.  In Section \ref{Sec:Production} we detail our method and workflow for the production of UV stabilized fibers through pre- and post-curing routes. In Section \ref{sec:Performance} we characterize the transmission performance of solarization-resistant fibers with respect to bend sensitivity, particle deposition and its mitigation, as well as their polarization-maintaining performance. We discuss our findings and conclude in Section \ref{sec:Conclusion}.




\section{UV-stabilized fiber patch cord production}\label{Sec:Production}

\begin{figure}[t]%
\graphicspath{{"Figure Source Material/Figure Loading/"}}
\includegraphics[width=\columnwidth]{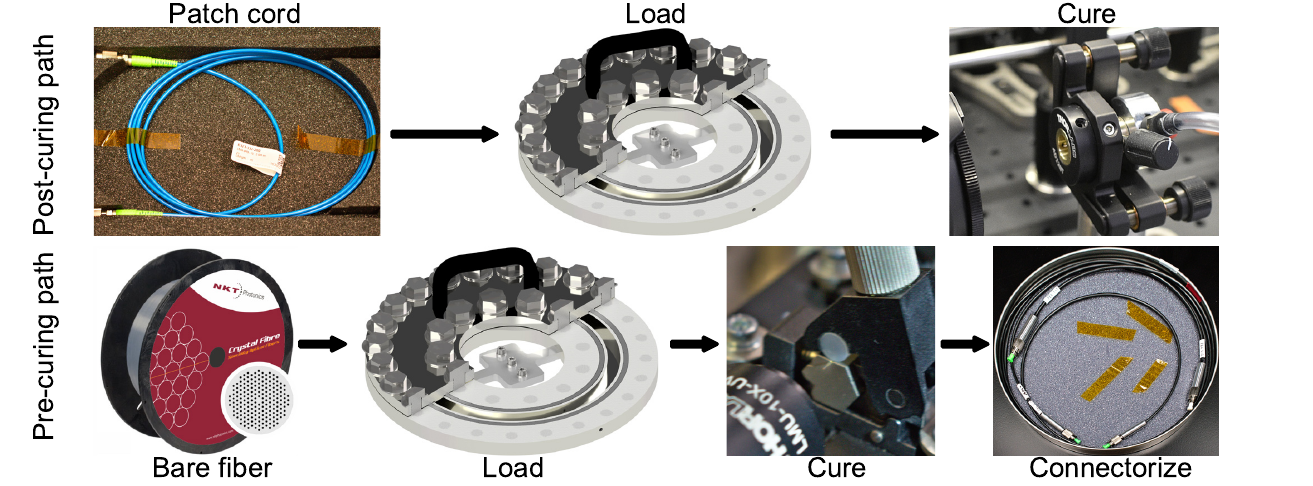}%
\caption{Pictorial representation of pathways to obtaining UV-compatible patch cords.}%
\label{fig:Loading}%
\end{figure}

We present two alternative techniques to obtaining the desired patch cords: the Post-curing and Pre-curing paths. The order of steps employed in the fabrication process distinguishes the two paths, and both present individual advantages. In the Post-curing path a commercially pre-connectorized fiber is loaded with \Ht\ in a suitable pressure vessel prior to curing by UV irradiation using standard patch cord optics. In the Pre-curing path bare fibers are loaded and then cured using standard bulk-optical fiber launching equipment before being commercially connectorized.  A summary of the processes is displayed in Fig.~\ref{fig:Loading}.

\subsection{Optical fiber selection and patch cord assemblies}
The fibers used in this manuscript are LMA-series photonic crystal fibers from \nkt, specifically LMA-10-UV and LMA-PM-10 with a $\unit[10]{\upmu m}$ mode field diameter. We select non-PM and PM fibers based on the competing objectives of retaining single-mode operation at short wavelengths, reduced flux densities in the core, and convenient handling. To understand this fiber choice, consider that step- or graded-index fibers achieve single-mode operation at short wavelengths by ensuring their normalized frequency or waveguide parameter $V$ satisfies

\begin{equation}
V = \frac{2\pi\rho}{\lambda}(n_\text{co}^2-n_\text{cl}^2)^{1/2} = \frac{2\pi\rho}{\lambda} \times NA \leq 2.405,
\label{eq:V_Standard}
\end{equation}

\noindent where $\rho$ and $\lambda$ are the fiber core radius and optical wavelength respectively, and $n_\text{co}$, $n_\text{cl}$ and $NA$ are the refractive index of the core and cladding and the fiber numerical aperture.  Traditional UV fibers therefore require cores sizes $\rho$ of only a few $\upmu$m in order to retain single-mode operation due to the inverse scaling in Eq.~\ref{eq:V_Standard} with wavelength.  In addition to requiring elaborate optics for proper mode matching, this incurs a quadratic increase in guided mode intensity per unit area.  This increased intensity both dramatically decreases the power levels which induce UV solarization and can also causes adverse nonlinear effects such as stimulated Brillouin scattering to occur at comparatively low power.

The use of photonic crystal fibers, by contrast, permits single-mode operation at reduced optical intensities. In one of the most common types of photonic crystal fibers, the cladding is defined by a equilateral-triangular air-hole pattern in place of a step or graded change in index, and the core is produced by omission of the central hole.  In this geometry the single-mode criterion is given by\cite{Nielsen2003a}:

\begin{equation}
V_\text{PCF} = \frac{2\pi\Lambda}{\lambda}\left(n_\text{co}^2(\lambda)-n_\text{cl}^2(\lambda)\right)^{1/2} = \frac{2\pi\Lambda}{\lambda} \times NA \leq \pi,
\label{eq:V_PCF}
\end{equation}

\noindent where $\Lambda$ is the hole pattern pitch and the refractive indices are now the strongly $\lambda$-dependent \textit{effective} indices of the fundamental and first eigenmode of the perfect, infinite cladding structure (for details see \cite{Nielsen2003} and references therein). This revised criterion\cite{Nielsen2003a,Nielsen2003} permits single-mode operation to be achieved with very large mode areas. The connection between $\Lambda$ and the mode area is only accessible numerically and studied in detail in \cite{Nielsen2003a} (see in particular Fig. 1 and 2). As we will discuss below, the hole pattern also has implications for \Ht\ loading.

Using these fibers our objective is to produce fiber patch cables with characteristics suitable for use outside of pristine laboratory environments run by expert users.  We produce fiber patch cables that are fully jacketed and employ an angle-polished finish on their connectors in order to mitigate unwanted etalon effects in particular during high-power operation.  

Axis-aligned, angle-polished and endcapped PCF patch cords are available from a number of vendors. The LMA-PM-10 based patch cords used in this study were \nkt' aeroGuide patch cords and custom connectorized patch cords from \nov. The aeroGuide patch cords have $\unit[3]{mm}$ plasticized PVC tubing and standard plastic boots to reinforce the connector region. Their connectors are zirconia ferrules from Diamond SA with a central stainless steel cap on the tip and active core alignment.  Endcaps are produced from bare fibers by tapering and then collapsing the hole pattern region. \nov-connectorized patch cords originally employed $\unit[900]{\upmu m}$ Hytrel jacketing without a reinforced boot. Later iterations also used $\unit[3]{mm}$ jacketing as well as reinforced stainless steel boots. Connectors on these cords are all zirconia and endcaps are added by splicing on pure silica segments. Both approaches have the benefit that the fiber tip sits flush with the connector, thereby reducing the potential to damage the facet, in contrast to self-connectorized patch cords as per \cite{Colombe2014} which leave the fragile fiber tip exposed. As with all end-capped fibers, traditional butt-coupling will incur additional insertion losses due to mode expansion in the non-guiding region.




\subsection{Hydrogen loading}
\label{sec:Diffusion}

The purpose of \Ht\ loading is to prevent the formation of active colour centers (other than E'-centers, also named group-V impurity-vacancy pairs, see \cite{Skuja2001} p. 157) by the incident (transmitted) radiation, or even to anneal pre-existing ones. The underlying physics of solarization and hydrogen annealing is discussed in detail elsewhere, but can be understood qualitatively by considering the following. A radiation field propagating through the silica core of a fiber provides the activation energy to break Si-O-Si bonds in the material that leave optically active colour centers. These colour centers absorb incident radiation and thus scatter light out of the guided mode (see \cite{Karlitschek1998,Karlitschek1998a} for discussion including reaction kinetics and \cite{Skuja2001} for an overview of different colour centers). 

If \Ht\ is present, it can bond with these colour centers via several different reaction pathways to form stable, optically passive states, thus imparting solarization resistance.  Of course, \Ht\ is not naturally abundant in silica fibers and therefore has to be introduced externally.  Common practice in both the UV curing and Bragg grating communities is to load by immersing the fibers in a \Ht\ environment for 2 weeks at $\unit[100-300]{bar}$ pressure. This process is accomplished by placing fibers or patch cords in pressure vessels and frequently heating the vessels in a water bath to both increase the maximum concentration achievable and reduce loading times. After loading fibers can be cooled for storage prior to curing.  To the best of the authors' knowledge there has been no quantitative study of requisite \Ht\ concentrations or a systematic comparison of performance under different loading conditions.  

A straightforward quantitative analysis helps us to identify loading conditions, including the in- and out-diffusion times germane to the curing and the influence of loading and storage temperatures on \Ht\ concentrations (target concentrations will be discussed later).  In the following we will outline briefly a model for hydrogen transport in silica to enable the user to identify loading conditions appropriate to their requirements. This model also informs relevant time scales for loading as well as out-diffusion, the latter serving as a limiting factor in storage times prior to curing.  

We model hydrogen transport in silica making two simplifying assumptions: the acrylic coating and plasticized PVC jacketing of any patch cord will not affect diffusion times and the presence of the hole structure in the fiber may be disregarded. Modeling the fiber as an infinitely long, uniform cylinder and imposing appropriate boundary conditions, we arrive at the specific solution for out-diffusion
\begin{equation}
C_\text{out}(r,t)=2\sum_{n=1}^{\infty}{\frac{J_0\left(\frac{\mu_n}{R_0}r\right)}{\mu_n J_1\left(\mu_n\right)}
\e{-D \left(\frac{\mu_n}{R_0}\right)^2 t},
}
\label{eq:FinalSolution}
\end{equation}
\noindent where we define the relative \Ht\ concentration $C_\text{out}(r,t)$ as a function of radial position $r$ and time $t$ (see Appendix for derivation). Here we use the $n$-th order Bessel function of the first kind, $J_n$, the fiber radius $R_0$, the diffusivity $D$ as a function of temperature $T$ and the $n$th node of the Bessel function $J_0$ is defined to be $\mu_n$. In- and out-diffusion are time-reversed processes and are therefore related by \mbox{$C_\text{out}(r,t)+C_\text{in}(r,t) = 1$}. 

With this quantitative model and an understanding of community protocols we can estimate required \Ht\ loading conditions and times for the photonic crystal fibers we employ. The inset in Fig.~\ref{fig:Curing} shows the solution at $r=0$ calculated for LMA-PM-10 parameters of $R_0=\unit[115]{\upmu m}$ and using \mbox{$D(T) = 2.83 \times 10^{-4} \exp\left(\frac{\unit[-40.19]{kJ/mol}}{N_\text{A} k T}\right)\unit[]{cm^2/s}$}\cite{Lemaire1991} with the Avogadro constant $N_\text{A}$ and the Boltzmann constant $k$. Diffusion times to reach a given relative concentration for a different fiber diameter can be obtained from the graph by noting from Eq.~\ref{eq:FinalSolution}

\begin{equation}
t_{x\%}(r_{1}) = t_{x\%}(R_{2})\times\left(\frac{R_1}{R_2}\right)^2,
\label{eq:RadiusAdjustment}
\end{equation}
where $R_2$ in the case of our graph is $\unit[115]{\upmu m}$. This relation holds exactly at the center of the fiber and is a good approximation as long as $r/R_0$ is small, \emph{e.g.} the actual concentration will differ by 5\% from the predicted if $r/R_0 = 0.2$.

In order to obtain information about absolute \Ht\ concentrations the relative concentration $C$ is multiplied by the solubility $S$ (in molecules per cm$^3$), which is the saturation concentration for a given pressure-temperature pair. For \Ht\ in silica the solubility is given by\cite{Shackelford1972}
\begin{equation}
S = p\left( \frac{h^2}{2 \pi m k T}\right)^{\frac{3}{2}}\frac{N_s}{kT}\left[\frac{\e{\frac{-\theta_\nu}{2T}}}{1-\e{\frac{-\theta_\nu}{T}}} \right]^3\e{-\frac{E_0}{N_\text{A}kT}},
\label{eq:Solubility}
\end{equation}
where $p$ is the pressure, $h$ is Planck's constant, $m$ is the mass of \Ht, $N_s = \unit[2.22\times10^{22}]{cm^{-3}}$ is the density of available lattice sites, $\theta_\nu=\unit[585.508]{K}$ 
is the characteristic temperature and $E_0 = \unit[-12.727872]{kJ}$ is the binding energy.

Calculations such as those above are particularly important in our study as we are considering the impact of fiber patch-cord configuration on \Ht\ diffusion time.  Recall that we made two assumptions in our model: no influence from the polymers nor from the air holes. The first is well justified as diffusivity and solubility of polymers for \Ht\ are all very similar \cite{Barth2013,SanMarchi2012} and transport speed through them is many orders of magnitude larger than through silica. The second however has a profound impact on the result.  

For \textbf{(i)}, an open LMA-PM-10, in- and out-diffusion times are 529 times faster than displayed in Fig. \ref{fig:Curing}, using Eq. \ref{eq:RadiusAdjustment}. This is due to the fact that the diffusivity of \Ht\ in air is $\sim10^{8}\times$ larger than in silica\cite{Marrero1972}, which means that once the hydrogen reaches these holes it can quickly diffuse out of the fiber, reducing the effective diameter to the diameter of the first cladding ring ($\unit[10]{\upmu m}$). Endcaps, if present, range in thickness between $\unit[50-100]{\upmu m}$, comparable to the fiber diameter. In lieu of numerical studies to model the complex diffusion dynamics including endcaps, a conservative estimate of the diffusion times can be made if the fiber is assumed to be \textbf{(ii)} solid (no holes) with its actual diameter for in-diffusion (thus overestimating the required time), and \textbf{(iii)} solid with thickness given by the minimum endcap thickness for out-diffusion (thus underestimating the storage time). 




\subsection{Practical considerations in \Ht\ loading and curing of bare and connectorized fibers}
\label{sec:Loading considerations}

\begin{figure}[b]%
\graphicspath{{"Figure Source Material/Figure Curing/"}}
\includegraphics[width=\columnwidth]{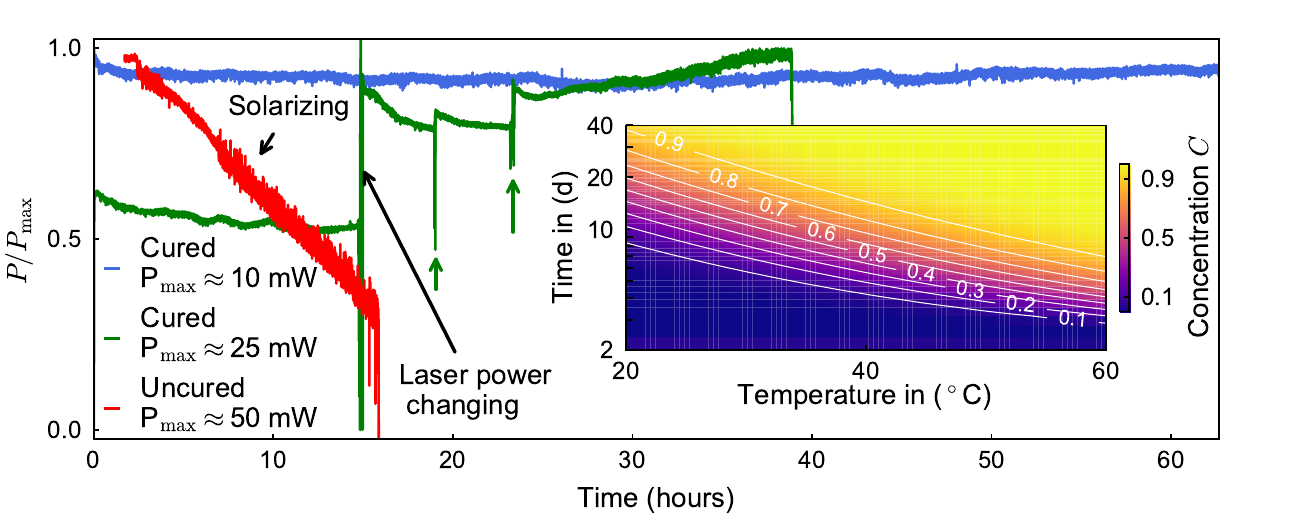}%
\caption{Normalized transmission through patch cords as a function of time. Fibers were a pre-connectorized LMA-PM-10 (blue, fabricated by \nov), an aeroGuide patch cord (green, fabricated by \nkt) and a pre-connectorized, unloaded LMA-PM-10 (red, fabricated by \nov). Green arrows indicate recoupling of the fiber after thermalization of the connector. Drifts in green curve dominated by drifts in input power. Inset: Relative concentration at $r=0$ as a function of temperature and diffusion time calculated for a $\unit[230]{\upmu m}$ diameter fiber. Colour and iso-concentration contours indicate change in concentration from starting value towards maximum/minimum value, e.g. for an empty fiber the 0.2 line indicates the time to reach 20\% full and for a saturated fiber it indicates 20\% lost. See text for details.}%
\label{fig:Curing}%
\end{figure}

In our experiments we procured several tens of meters of bare LMA-PM-10 and LMA-10-UV as well as patch cables of different lengths between $\unit[1.8 - 5.0]{m}$, and embarked on loading and curing all fibers.  However, there are important practical differences in loading bare and connectorized fibers that we discuss in context here.  Barring quantitative studies on required concentration, a guide to what target concentration, $n_\text{target}$, to use is given by the absolute concentration values for common practice in the community. From the Figs. presented above we deduce a concentration at the center of the core at the start of curing at around $\unit[(1\times10^{20}-1\times10^{22})]{cm^{-3}}$, depending on pressure, fiber diameter and time between loading and curing.  From Fig.~\ref{fig:Curing} one can now easily find the required time to load to a desired concentration of \Ht\ in the fiber for any of these cases and any cladding diameter in four steps:
\begin{enumerate}
	\item Determine target concentration $n_\text{target}$
	\item Calculate fraction of solubility $S(p,T)$, i.e. $x\% = n_\text{target}/S(p,T)$
	\item Find time $t$ for temperature $T$ that reads $x\%$
	\item Adjust time as per Eq. \ref{eq:RadiusAdjustment} for the three cases above.
\end{enumerate}

Loading times to a given concentration can be decreased dramatically by choosing pressure and temperature pairs correctly. For example, the loading time is decreased by a factor of 7.24, with unchanged solubility, by moving from $\unit[100]{bar}$ at $\unit[20]{^\circ C}$ to $\unit[160]{bar}$ at $\unit[60]{^\circ C}$. This is due to the solubility increasing linearly with pressure and decreasing slowly, but in a complex fashion, with temperature in the applicable region. 

We heat our pressure vessel via immersion in a temperature-controlled water bath during loading of connectorized fibers to $\unit[60]{^\circ C}$ to compensate for the increase in in-diffusion time by the presence of the endcaps [refer to Fig. \ref{fig:Curing}]. Fiber patch cords should not be exposed to temperatures well above $\unit[60]{^\circ C}$ to prevent melting or embrittlement of the polymers and glues in use\cite{CatoCommunication}.  Loading connectorized patch cords has not, in our studies, presented any problems from \Ht-induced material expansion. 


Following hydrogen loading, optical fiber patch cords must be UV cured before out-diffusion of the \Ht\ from the fiber. The timescale for this process is determined from the quantitative analysis above, but we typically commence curing within 24 hours of fiber removal from the loading vessel for endcapped fibers. In bare fibers, one can take advantage of the fast diffusion through the hole structure by loading an open PCF and quickly collapsing the ends after removal from the loading vessel. This approach allows users to benefit from the enhanced loading speed of open fibers as well as the increased storage time of collapsed fibers (It is important to note that \Ht\ in a fusion splicer can cause audible micro-explosions which, while not dangerous, can lead to the formation of bubbles in the collapsed region.).  Using endcapped fibers, out-diffusion times will be substantially increased from hours to days or, if cooled, months and years. We have on occasion employed cooling to $\unit[-70]{^\circ C}$ in order to inhibit out-diffusion and extend hold times of loaded fibers.  If open-ended fibers are to be cured, treatment must occur shortly after removal from the vessel because out-diffusion times at room-temperature are on the scale of few hours and curing itself takes many hours. 

Significant practical considerations again emerge in UV curing for the two fiber production paths, as was the case with \Ht\ loading.  Curing is experimentally achieved by coupling $\unit[313]{nm}$ laser light produced in a two-step, nonlinear frequency conversion system~\cite{Wilson2011} into the fibers.  Input power ranges from $\unit[10-200]{mW}$, and as detailed below achieves comparable final performance in terms of solarization robustness irrespective of curing power.  For the post-curing path conventional fiber collimators and refractive optics simplify coupling dramatically, making this an attractive option for many laboratories. 

Considerably more care must be taken when attempting to couple into bare fibers in the pre-cured path. In this case bare fibers must be properly cleaved and placed in a suitable fiber launching setup constructed from bulk optics. Achieving good coupling can initially be very challenging, so reducing the number of degrees of freedom by e.g. mounting the launcher and lenses rigidly in a cage system has proven very helpful. Bending the fiber can make finding the fundamental rather than the speckle mode easier. The large MFD of the LMA fibers means that plano-spherical lenses and, depending on input mode quality, a microscope objective, are sufficient to achieve maximum coupling efficiencies. When working with open fibers, pre-aligning the setup using an unloaded fiber section is advisable to minimize delays in curing.  For further advice on bare fiber coupling in the context of UV curing see also the NIST fiber recipe\cite{Recipe} (in particular pages 10 and 11).



\section{Performance characterization of solarization-resistant fibers}
\label{sec:Performance}
In our experiments we have characterized five pre-cured and two post-cured fibers guided by the insights introduced above. In Fig.~\ref{fig:Curing} we present transmission logs during the initial curing process for three different patch cords. Two of these were \Ht\ loaded and cured during this irradiation, while a third unloaded fiber was used as a control for comparison. In all cases the cumulative optical energy $E_\text{tot}=\int_0^t{P(t')\text{d}t'}$ at the output is $\geq \unit[1.8]{kJ}$, achieved in practice by using $\sim\unit[20]{mW}$ measured at the fiber output over a $\unit[24]{h}$ curing period.

In our measurements we do not observe any measurable degradation in transmission across a range of input powers for \Ht\ loaded fibers. This performance is maintained over long periods without further need for hydrogen loading; for instance, the patch cord used to acquire data represented in green was cured seven months prior to preparation of this manuscript and used in a variety of other measurements. Data from the same fiber is shown in Figs. \ref{fig:BendingLoss}, \ref{fig:Flushing} and \ref{fig:Polarization}. The fiber has accumulated E$_\text{tot} > \unit[10]{kJ}$ with no measurable reduction in coupling efficiency or transmission.  By contrast, the unloaded fiber exhibits a steady decay in transmission over the timescale of hours that is irreversible.  A key signature of solarization is that the residual optical transmission is dominated by the speckle mode, while transmission in the fundamental is negligible.  We have verified that the higher input power employed in the fiber that solarized here is not linked to the onset of solarization, as other fibers have been successfully cured using powers up to $\unit[200]{mW}$ and cumulative energy is identical or larger for cured fibers.

In our tests, bulk transmission through cured fibers typically is between 67\% and 75\%, with one cable at 80\%. We have not produced sufficient patch cables at this stage to clearly distinguish bulk transmission losses vs coupling efficiency.  Despite that, we have some evidence that variation in the connectorization process can contribute to the measured transmission; while most patch cords show symmetric transmission from both connectors, two \nov\ pre-cured fibers with steel-reinforced boot show asymmetry (50\% vs 70\%) indicative of variations in connector quality between sides.  While at $\unit[313]{nm}$ the LMA-PM-10 is technically no longer single-mode, we have only observed higher-order fiber modes when higher-order modes from our laser source were incident on the fiber.

\subsection{Mitigation of bending sensitivity in UV transmission}
Bend sensitivity of photonic crystal fibers can reduce optical transmission to zero in both cured and uncured fibers and is therefore an important phenomenon to consider in both patch cord curing and subsequent handling of the finished fibers.  This sensitivity becomes problematic in the UV as PCFs possess both an upper and lower wavelength sensitivity edge~\cite{Birks1997}, in contrast to step-index fibers which only have an upper wavelength edge.  The critical short-wavelength bending radius for PCF\cite{Birks1997,Nielsen2004} is \mbox{$R_c\propto \Lambda^3/\lambda^2$}, and accordingly the use of a larger fiber core (and hence larger $\Lambda$) increases bend sensitivity. 

\begin{figure}[b]%
\graphicspath{{"Figure Source Material/Figure Bending/"}}
\includegraphics[width=\columnwidth]{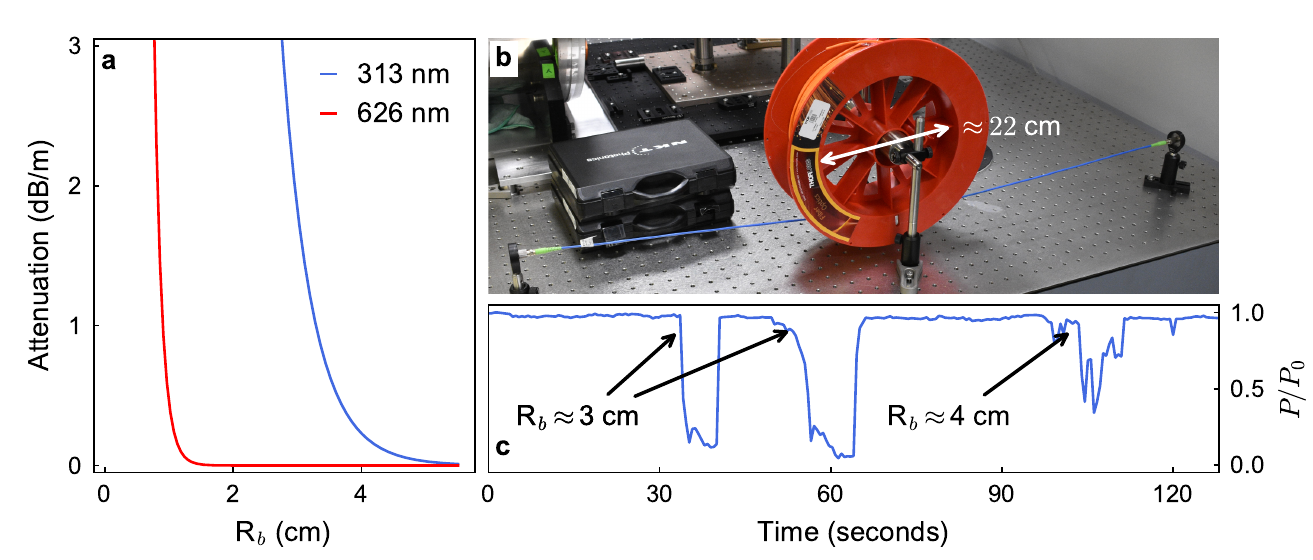}%
\caption{(a): Calculated bending losses for an LMA-PM-10 based patch cord as a function of bending radius for two wavelengths. Calculation based on \cite{Nielsen2004} with the following parameters for LMA-PM-10: lattice pitch $\Lambda = \unit[6]{\upmu m}$, refractive index of silica n$_s = 1.444$, and hole diameter $d = \unit[3]{\upmu m}$, which inform the values for V$^* \approx 3.75$ and A$_\text{eff} \approx \unit[36.75]{\upmu m^2}$ from references \cite{Nielsen2003,Nielsen2003a}. (b): Patch cord on stretching jig to reduce bend losses after loading. Fiber connectors are fixed to Thorlabs SM1FCA on steel pillars rotating freely in their posts such that the fiber does not bend at the boot. Patch cords are kept under sufficient tension to allow for strongly damped string oscillation so the jacket is straightened. (c): Fiber transmission as function of time with sharp bends induced. First two drops induced by sharp bends at intermediate positions along the fiber. The last drop caused by a bend induced at the boot. 
}%
\label{fig:BendingLoss}%
\end{figure}

Figure \ref{fig:BendingLoss}(a) shows the predicted macrobending loss \cite{Nielsen2004} as a function of bending radius in a LMA-PM-10 fiber for two wavelengths used in our measurements, $\unit[626]{nm}$ and $\unit[313]{nm}$, clearly demonstrating the wavelength dependence of this phenomenon.  The impact of experimental application of tight-radii bends in the fiber on optical transmission in the UV is shown in Fig.~\ref{fig:BendingLoss}(c). Transmission can be reduced to $\sim50\%$ through application of bends with radii $\mathbf{R}_{b}\approx \unit[4]{cm}$ and to $\sim10\%$ for $\mathbf{R}_{b}\approx \unit[3]{cm}$ identical to what was found for bare fibers in \cite{Colombe2014}.

A key source of local bending losses in the production of commercial UV-compatible fiber patch cords arises from plastic memory in the jacket. This can be sufficiently severe as to inhibit transmission and ultimately even impede curing.  In our experiments this phenomenon is exacerbated by the fact that the loading vessel we use possesses a relatively tight $\unit[10]{cm}$ diameter.  In addition, the use of elevated temperature during loading makes the jacketing more malleable and introduces memory after returning to room temperature. It is not known whether the presence of high-pressure \Ht\ contributes to this phenomenology, but we note that similar effects have emerged in coiled fibers after sitting in atmosphere post curing.

We overcome this issue by stretching the jacketed fibers after loading and during the curing procedure using a simple stretching jig [Fig.~\ref{fig:BendingLoss}(b)].   We used commercially available plastic reels with $\unit[22]{cm}$ diameter during the curing process, and $\unit[45]{cm}$ diameter reels for storage.  Our measurements demonstrate that stretching for 1-2 weeks offsets bending losses as the fiber jacket relaxes, and following jacket relaxation (and successful curing) we do not observe significant transmission sensitivity to fiber orientation. A properly stretched patch cord or bare fiber will show only a few percent transmission variation if disturbed mechanically, but fixing patch cords and fibers in position remains advisable for any sensitive task.  

\begin{figure}[h!]%
\graphicspath{{"Figure Source Material/Figure Vessel/"}}
\includegraphics[width=\columnwidth]{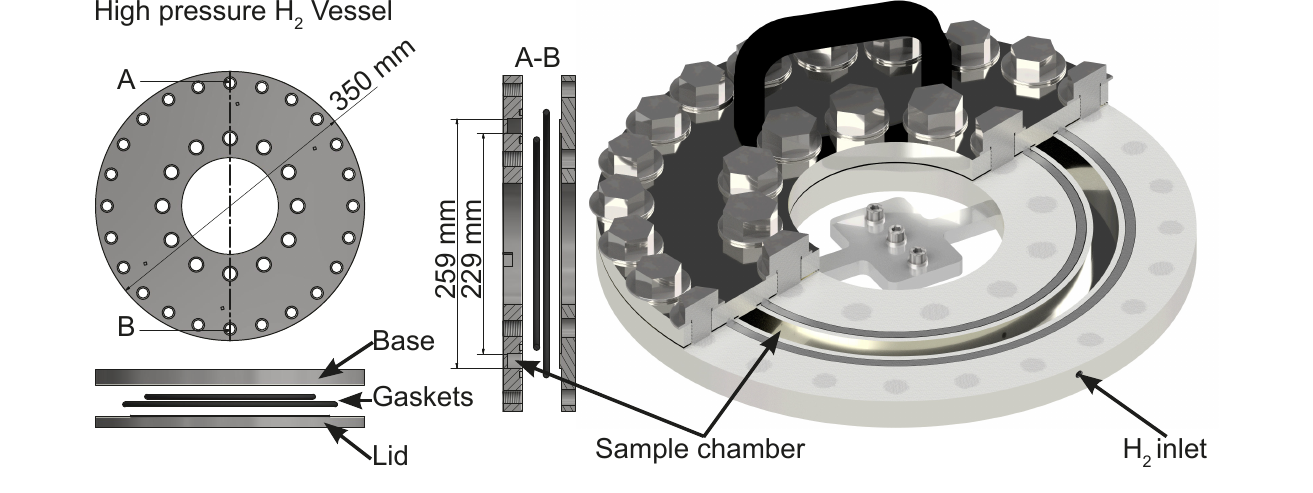}%
\caption{High pressure H$_2$ vessel design for loading fiber patch cords with mitigated bend losses through plastic jacket memory.  Fibers sit in the pressurized low-volume channel denoted ``sample chamber.'' The vessel is pressurized with \Ht\ from a manifold via a 1/4"-NPT tube fitting on the \Ht\ inlet. Fastening (tapped holes of 5/8-11 UNC - 2B and 3/4-10 UNC - 2B with corresponding hex head bolts of 5/8x1.25 and 3/4x1.25 from SA-320 B8 Class 2) and plate thicknesses selected on advice following design validation and compliance assurance conducted in cooperation with an accredited engineering consulting firm. Left: Top-down view, explosion view of parts and intersection view from A to B along indicated line. Right: Ray traced image of validated design.}%
\label{fig:H2_Vessel}%
\end{figure}

These observations help address a tradeoff we experienced in our design process.  We observe that fibers with thin jacketing exhibit lower memory-induced bending loss, but ultimately provide low mechanical protection and remain sensitive to vibration and mechanical perturbation.  On the other hand, fibers using thick jacketing are ultimately more mechanically stable but initially present with significant plastic memory, complicating fiber coupling and curing.  The synthesis of our results is that a suitably stretched fiber with thick jacketing is preferred, and that the radius of curvature of the high-pressure loading vessel should be maximized.  Accordingly, we present the design of an optimized pressure vessel in Fig. \ref{fig:H2_Vessel}, which we are currently in the process of constructing.  This design provides sufficient sample chamber diameter and width to allow the entire boot section to remain straight during loading, as the sharpest bends occur at this point in small vessels.  Considering aeroGuide patch cords we select an outer diameter $\sim\unit[260]{mm}$ and the chamber is constructed as a groove in order to minimize the requisite volume of hydrogen in the loading procedure. Rubber gaskets are used to seal the sample chamber in the center of the annulus. 




\begin{figure}[p]%
\graphicspath{{"Figure Source Material/Figure Flushing/"}}
\includegraphics[width=\columnwidth]{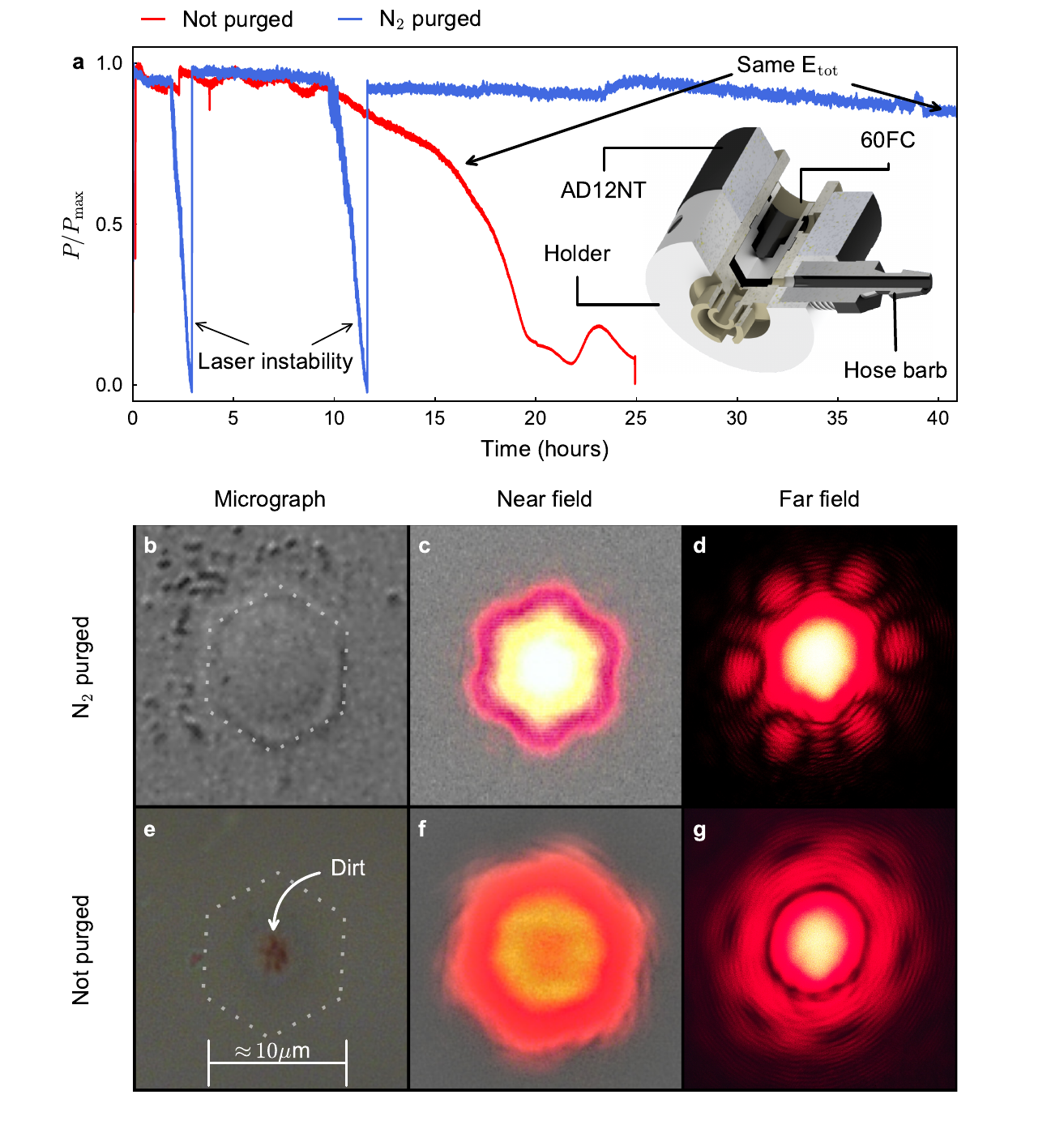}%
\caption{(a): Normalized transmission through LMA-PM-10 patch cords as a function of time. Decreasing transmission due to scattering-center deposition on the fiber facet (red curve) is reversible with polishing.  Purging the fiber collimator with dry N$_2$ mitigates this phenomenon. Dips in power stem from laser instabilities. Inset shows a mounting assembly for N$_2$-purged 60FC collimators from Sch\"after + Kirchhoff. The collimator is mounted in the front using a Thorlabs AD12NT adapter for 1'' optics. The flushing jig is aligned to position the hose barb above a venting hole drilled in the space between lens tube and fiber facet. Two set screws fasten the jig. Lens adjustment with jig in place is possible on longer focal length collimators. (b): Micrograph of a N$_2$-flushed aeroGuide fiber facet (guided fundamental mode at 633 nm), showing magnified $\approx \unit[10]{\upmu m}$ diameter core region where the hexagonal mode shape is visible in the glass after many hours of UV transmission with no detrimental effects on coupling efficiency. This pattern is not the actual hole pattern as this image is focused on the facet rather than the hole structure behind the endcap. (c): Near-field transmitted light using a fiber alignment tool, imaged with the detector in saturation in order to show the fundamental (Gaussian) and the hexagonal pattern arising from the hole pattern which is invisible without saturating the detector.  (d): Far field transmission.  (e)-(g): Same as for (b)-(d) above, but for a connector which was not \Nt-purged. Polishing the deposits away restores the observations in (b)-(d).}%
\label{fig:Flushing}%
\end{figure}

\subsection{Facet contamination and \Nt-purged collimators}
\label{sec:Purging}

The deposition of particles or other contaminants on the fiber facets [Fig. \ref{fig:Flushing}(e)] is another UV-exacerbated mechanism that can compromise fiber performance. Firstly, in this wavelength regime the optical-tweezer effect for dielectric particles is large due to high optical intensity in the relatively small mode diameter.  Secondly, light absorption by particles and aerosols (thus burning them into surfaces) can become important as most common particles such as dust or oil in air will have strong transitions in the UV but not in the visible or IR. In either case, scattering centers form on facets which deteriorate mode matching or compromise output mode quality, even during the curing process. In Fig.~\ref{fig:Flushing}(a) the deterioration of transmission through a fiber is shown (red curve) when UV is incident.  In contrast to the process of bulk fiber solarization, transmission can be restored to previous levels simply by polishing the fiber facets.  

We mitigate this effect by actively purging fiber collimators with low-flow, clean, dry \Nt.  The purging process is accomplished using a modified 60FC collimator from Sch\"after + Kirchhoff [inset Fig.~\ref{fig:Flushing}{a}], and low-flow \Nt\ obtained from a building-wide system recovering liquid boiloff is applied using a hose barb on the collimator. In our system we flow $\unit[(0.5 - 1.0)]{sL/min}$, but the required flow rate will depend on type of tubing used, leak rate, fit of the jig and so on. The general aim here is to provide minimal overpressure in the volume between fiber and collimator lens to prevent particle migration into this region. The blue curve in Fig.~\ref{fig:Flushing}(a) represents an example measurement in which this \Nt\ purge prevents the onset of facet degradation. The viability of the approach has been demonstrated over several months, where without purging we have seen deposition of particles on fibers with input powers as low as $\unit[10]{mW}$. In addition, we observed asymmetric contamination behavior on the input/output facets of a fiber where each end was purged/ not-purged respectively. 

Visual inspection of the fiber facet and the output mode both using and omitting \Nt\ purging demonstrates the efficacy of this approach. Figures \ref{fig:Flushing}(b)-(d) show a micrograph of a fiber endcap of a clean, purged patch cord along with the near- and far-field of the same fiber.  In Fig. \ref{fig:Flushing}(c) transmission through the fiber in the near field reveals the approximately Gaussian fundamental mode, as well as the hexagonal pattern in the output mode arising from the hole structure. In the far-field we observe the Fourier transform which looks qualitatively similar [Fig. \ref{fig:Flushing}(d)].

When collimators are not purged and contaminants are deposited on the fiber facet, the impact on the transmitted mode is significant.  Figs.~\ref{fig:Flushing}(f)-(g) show similar images for a collimator that was not purged, resulting in visible contamination buildup [Fig.~\ref{fig:Flushing}(e)]. In the near field a minimum is visible in the center of the fiber due to the accumulated contamination. The far-field pattern in Fig. ~\ref{fig:Flushing}(g) deviates more strongly from the expected near-Gaussian profile, has a much higher fraction of energy distributed outside the central region, and rapidly diverges spatially. These observations can be contrasted with observations under solarization, where an increasing fraction of light appears in the speckle mode image of the fiber (including stress rods and dark hole structure). Characterization of the far-field output mode therefore provides a useful diagnostic to distinguish between various sources of transmission loss.




\subsection{Characterization of polarization-maintaining performance}
\label{sec:Polarization}
\begin{figure}[b]%
\graphicspath{{"Figure Source Material/Figure Polarization/"}}
\includegraphics[width=\columnwidth]{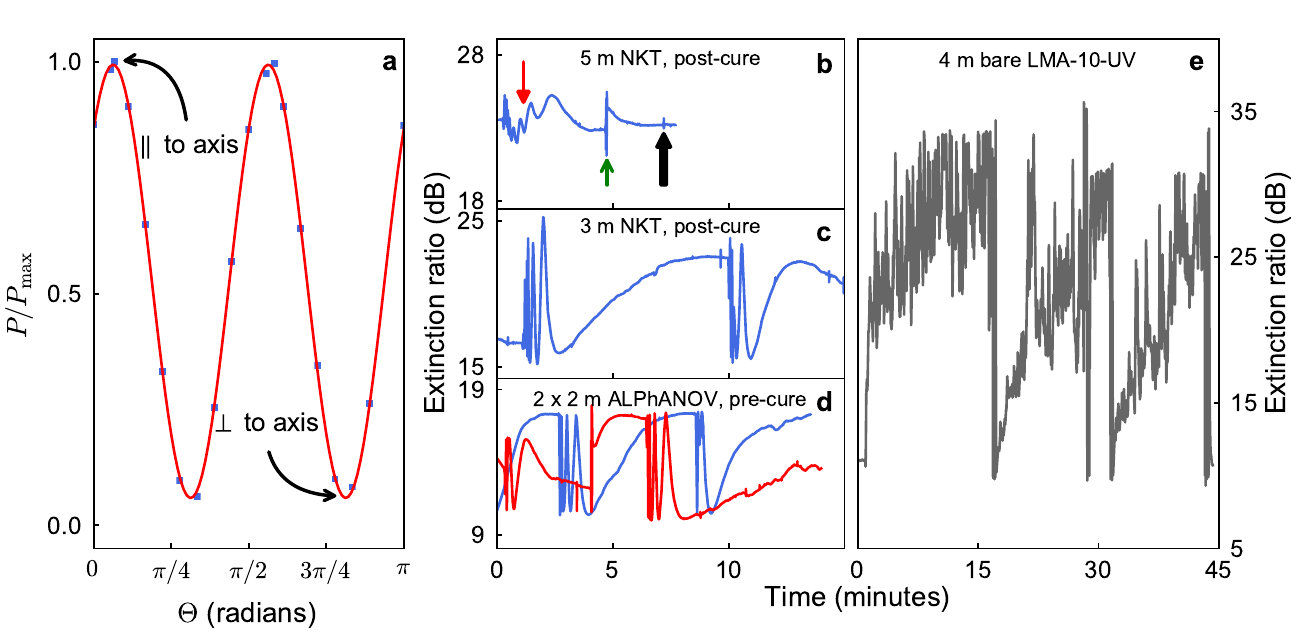}%
\caption{(a): Representative data for normalized transmission through UV patch cord as a function of input polarization angle. Data taken with $\unit[5]{m}$ aeroGuide patch cord at $P\approx\unit[30]{mW}$. Axis refers to the fiber slow axis. (b)-(d): Polarization extinction ratios measured behind LMA-PM-10 fiber based patch cords. Panel (d) has two different fibers of the same kind and length. Mechanical or thermal strain is applied to modify birefringence. Arrows indicate typical signatures of thermal strain (red), mechanical strain (green), and artifacts from laser lock disturbances (black). (e): Polarization extinction ratios measured behind LMA-10-UV bare fiber, again under themral and mechanical stress. Note the different time scale relative to panels (b)-(d).  All PER measurements used $P\approx\unit[10]{mW}$ at the fiber output.}%
\label{fig:Polarization}%
\end{figure}

A critical demonstration in this study is that PM fiber patch cords may be produced as a complement to previous non-PM fibers.  We characterize fibers produced under both pathways outlined above using a polarization analysis setup.  The primary initial observation is that, although unintended, all fibers behave as linear polarizers after curing. The degree of polarization we have observed varies between fibers and can range from $\sim15\%$ to near full polarization.  The origin of this observation in our UV fiber patch cords is not understood, although such phenomenology is well known and sometimes specifically designed in fibers used at longer wavelengths~\cite{Schreiber2005}. Figure~\ref{fig:Polarization}(a) shows typical normalized transmission through a $\unit[5]{m}$ long LMA-PM-10 based aeroGuide patch cord as the input polarization angle is varied using a $\lambda/2$ waveplate. Similar behaviour is observed for all tested powers and in all tested patch cords, including \nov-connectorized, pre-cured patch cords. Since the transmission maxima occur when the input polarization is aligned with the slow axis, optimizing input polarization alignment becomes equivalent to maximizing transmission. 

We measure the polarization extinction ratio (PER) using a Wollaston polarizer to analyze light transmitted through a fiber. Such measurements for PM fibers are shown in Fig.~\ref{fig:Polarization}(b)-(d), where fibers were strained either mechanically or thermally to modify the birefringence in situ. All PM patch cords display qualitatively similar behavior, with PER values in the range of 15-25 dB, with unperturbed variations $\sim \unit[0.1]{dB}$, but \nov-connectorized, pre-cured patch cords perform somewhat less favorably amongst the samples we have on hand.  All patch cords show moderate reactions to thermal and mechanical strain with typical deviations of order $\unit[5-8]{dB}$ in response to strong disturbances, {\emph e.g.} application of heavy pressure or a strong, $\unit[70]{^\circ C}$ air flow, but show no significant response to minor thermal or mechanical disturbances typically encountered in the laboratory. We observe that residual strain and bending of the fibers induced coupling of light into the unpolarized speckle mode, producing a limit on measured PER. Fiber performance returns to unperturbed levels within several minutes.

In addition to the measurements above, all PM patch cords were also characterized at $\unit[369]{nm}$ using a Sch\"after + Kirchhoff polarization analyzer (SK010PA-UV).  Under these conditions, measured PERs were recorded between $\unit[23-26]{dB}$ (pre-cured patch cords) and $\unit[30-32]{dB}$ (post-cured patch cords). These values can be compared to e.g. the guaranteed value of $\unit[15]{dB}$ at $\unit[405]{nm}$ for Thorlabs' PM-S405-XP.

For comparison we present comparable measurements on a bare, non-PM $\sim\unit[4]{m}$ long LMA-10-UV fiber in the same $\unit[313]{nm}$ measurement setup in Fig.~\ref{fig:Polarization}(e), except for the required use of appropriate coupling optics. The fiber was cleaved on both ends and the output mode was inspected to ensure good performance. The unperturbed PER was measured $\unit[30-35]{dB}$ with unperturbed variation of order $\unit[5]{dB}$, likely limited by the measurement setup.  However this varied strongly between fibers tested with another sample showing PER of only $\unit[3-5]{dB}$. We believe these differences may arise due to variation in the quality of the cleave; in the low-PER sample a faint scatter ring around the output was perceptible, though the output spatial mode showed no disturbance. 

The bare fiber shows significantly higher sensitivity to external influences than the patch cords, with peak-to-peak variation in PER $\sim\unit[25]{dB}$ in response to mechanical or thermal strain. Recovery times under application of thermal strain are also much longer, taking $\sim\unit[15]{min}$ to stabilize, while recovery from mechanical strain requires only a few seconds. Bare fibers show sensitivity to minor external disturbances as occur during normal laboratory work. Contrary to patch cords there is no detectable speckle mode contamination in the PER measurements and the remaining power in the undesired port can be easily identified as the orthogonally polarized fiber mode.

\section{Discussion and conclusion}
\label{sec:Conclusion}
In this work we have presented two pathways toward fully commercial, UV-compatible fiber patch cords suitable for a wide range of demanding applications. Patch cords were produced following both pathways and their transmission and polarization-preserving properties characterized. We successfully loaded and cured open-ended (not endcapped) bare fibers as well as endcapped and pre-connectorized fibers with typical bulk transmission between $\unit[66-75]{\%}$. While the endcaps themselves will not be cured after connectorization on the pre-curing path, we have found no detectable drop in transmission through their solarization. Experiments conducted on bare fibers suggests this observation may not hold for deeper UV applications~\cite{ColombeCommunication}, where endcaps showed strong solarization on exposure to $\unit[267]{nm}$ light.  In this case, incident photons overlap with the absorption peak of two prominent colour centers (bulk interstitial O$_3$ and NBOHO centers\cite{Skuja1998}); for such applications endcaps will need to be re-cured even on the pre-curing path.

Measured polarization extinction ratios for our patch cords range between $\unit[10-25]{dB}$ for all patch cords and $\unit[15-25]{dB}$ for the post-cured ones. Our PER measurements on PM UV patch cords and bare non-PM fibers at $\unit[313]{nm}$ convincingly demonstrate that jacketing and connectorization shield against external influences in addition to providing safer and more convenient handling.  However our measurements also reveal the connectorization and cleave quality have a significant influence on performance. This again highlights the particular challenges when working with bare fibers because visual inspection of the facet or the output mode may provide insufficient to determine acceptable performance. The influence of the connectorization quality also means that comparison between the absolute value of the PER for bare fibers and patch cords is very difficult, although their dynamic ranges in response to external perturbations may be compared. Since vendors currently tend to lack equipment to perform tests at low wavelengths quality assurance remains challenging and some scatter in the achievable PERs is to be expected. It is at this point not clear whether the superior performance of our aeroGuide post-cured patch cords is due to the more mature connectorization process or due to an intrinsic difference in the two pathways, although the former seems most likely.

Interestingly, neither curing nor solarization had any measurable effect on the fiber performance at $\unit[626]{nm}$ in mode shape, coupling efficiency, polarization maintenance or bend sensitivity.  At $\unit[626]{nm}$ overall performance specifications were observed to be qualitatively similar to those presented above. As expected, however, visible wavelength performance of PER and bend sensitivity is superior owing to the longer wavelength. PER values measured at $\unit[626]{nm}$ using a polarimeter were between $\unit[30-55]{dB}$ and bend sensitivity is almost completely negligible.

Other groups \cite{Colombe2014} have cured at significantly higher peak power or identically $E_\text{tot}$, while the data in \cite{Karlitschek1998} shows successful curing at as little as $\unit[0.5]{J}$ using a pulsed laser source.  We believe that the use of endcapped fibers assists in the efficacy of low-power curing as the \Ht\ concentration does not change appreciably through diffusion during the curing process at room temperature. In addition there is no need for repeated recoupling caused by fiber/connector thermalization. A change in coupling can nonetheless often be observed which is conjectured to stem from a change in refractive index through the solarization as well as annealing of pre-existing colour centers\cite{Karlitschek1998}.

As the authors of \cite{Karlitschek1998,Karlitschek1998a} point out the OH content of the glass used in the fiber may be of importance even when \Ht\ loading and UV curing is employed. As such these two processes can be beneficially combined for two main reasons. First, the glass background absorption after curing high-OH content fibers is generally lower than with low-OH content fibers. Secondly, some spectral features are not present in high-OH content glass after successful treatment where they remain in low OH glass. Unfortunately, high-OH PM fibers are currently unavailable, but are under development at \nkt.

It is clear that solarization resistance is imparted for wavelengths other than that employed in the curing process, which can at least in part be attributed to the fact that homogenous broadening of colour centers is typically much larger than inhomogeneous broadening \cite{Skuja1998}. For instance, it was demonstrated that curing at $\unit[375]{nm}$ with $\unit[50]{mW}$ imparted solarization resistance for $\unit[313]{nm}$ and $\unit[200]{mW}$, and curing at $\unit[313]{nm}$ with $\unit[200]{mW}$ imparted solarization resistance for $\unit[235]{nm}$ and $\unit[10]{mW}$\cite{WilsonCommunication}.  However there is to our knowledge no quantitative study characterizing the breadth of the solarization-resistance window, or  power handling characteristics for a given curing power. This remains an important area of future study.

While both production pathways are ultimately viable, we identify that the pre-curing path may be most compatible with large-scale production of fiber patch cables for near-UV application, where endcap solarization does not appear problematic. Post-curing reduces the time required to produce small numbers of fibers, but mandates the fabrication of a larger loading vessel in order to minimize bending losses and multiple curing runs. Long-sections of bare fiber may be loaded and cured in a single shot before being sent off to be cut to size, connectorized, and jacketed.  We hope that these results present the community with a useful tool for the next generation of experiments in diverse fields and provide an important step towards a high-quality, fully commercial solution.

\section*{Appendix I: Derivation of solution for diffusion equation}

Hydrogen transport in a uniform medium is governed by a diffusion equation of the following form:

\begin{align}
\partial_t C\left(\vec{r},t\right) &= D(T)\Delta C\left(\vec{r},t\right)\label{eq:DiffusionEquation1}\\
\partial_t C\left(r,t\right) &= D(T)\left(\partial_r^2 C\left(r,t\right)+\frac{1}{r}\partial_r C\left(r,t\right)\right) = \frac{D}{r}\partial_r\left[r\partial_r C\left(r,t\right)\right],
\label{eq:DiffusionEquation2}
\end{align}
\noindent where we denote the (relative) concentration as a function of position $\vec{r}$ and time $t$ as $C$, $D(T)$ is the temperature-dependent diffusivity and $\Delta$ is the Laplace operator. From Eq. \ref{eq:DiffusionEquation1} to \ref{eq:DiffusionEquation2} we have moved into cylindrical coordinates such that $r = \left\|\vec{r}\right\|$ and we have assumed both axial symmetry and an infinitely long cylinder. Assuming the last is a good approximation since the fiber length will be several orders of magnitude larger than its width so any variation with $z$ will only impact the ends (Variation at the ends will be dominated by the air hole pattern. The boundary conditions for the radial and longitudinal part are coupled because of this hole pattern, so numerical studies to explore the complex dynamics arising would be necessary).

This equation can be solved using the separation of variables approach, which leads to a first order ordinary differential equation in time and Bessel's differential equation of order 0 in radial position. The general solution is given by
\begin{equation}
C\left(r,t\right)=A_1\left[A_2 J_0\left(\lambda r\right) + A_3 Y_0\left(\lambda r\right)\right]\e{-\lambda^2 D t},
\label{eq:GeneralSolution}
\end{equation}
\noindent where the $A_i$ are free parameters, $-\lambda^2$ is the separation constant and $J_0$ and $Y_0$ are the Bessel functions of the first and second kind of order 0, respectively. To obtain the specific solution we impose the physical requirement for the concentration to remain finite at the center at all times ($A_3 = 0$) and (for out-diffusion) require the concentration on the boundary to vanish at all times $C_\text{out}(r=R_0,t)=0$, i.e. the fiber out-diffuses into a well ventilated area. The boundary condition yields

\begin{equation}
J_0(\lambda R_0)\stackrel{!}{=}0 \Rightarrow \mu_n = \lambda_n R_0 \Leftrightarrow \lambda_n = \frac{\mu_n}{R_0},
\label{eq:BesselZeros}
\end{equation}
\noindent where $\mu_n$ is the $n$th zero of $J_0$ and the index $n$ counts through all independent solutions. Since there are now an infinite number of linearly independent solutions, the most general form of our specific solution is
\begin{equation}
C\left(r,t\right)=\sum_{n=1}^{\infty}{A_n J_0\left(\frac{\mu_n}{R_0}r\right)\e{-D \left(\frac{\mu_n}{R_0}\right)^2 t}}.
\label{eq:nSolution}
\end{equation}

\noindent To fix the last free parameter $A_n = A_{1,n} A_{2,n}$, with $A_{i,n}$ being the $n$th free parameter $A_i$ from Eq. \ref{eq:GeneralSolution}, we substitute the initial condition, which leads to
\[C(r,0)=C_0(r)=\sum_{n=1}^{\infty}{A_n J_0\left(\frac{\mu_n}{R_0}r\right)},\]
which can be recognized as the Fourier-Bessel representation of $C_0(r)$. In the usual manner the series coefficients are found by projection and normalization.
\begin{equation}
A_n=\frac{\left\langle C_0(r)|J_0\left(\frac{\mu_n}{R_0}r\right)\right\rangle }{\left\langle J_0\left(\frac{\mu_n}{R_0}r\right)|J_0\left(\frac{\mu_n}{R_0}r\right)\right\rangle }=2 \frac{ \int_0^{R_0}{r' C_0(r')J_0\left(\frac{\mu_n}{R_0}r'\right)\text{d}r'}}{R_0^2J_1\left(\mu_n\right) }.
\label{eq:FourierBessel}
\end{equation}
\noindent Here $\left\langle |\right\rangle$ denotes the scalar product over the required metric (in the case of the Bessel function the weight is $r$) and the denominator was calculated using the orthogonality relation of the Bessel functions.

In principle the general solution is now found, as all free parameters have been set,
\begin{equation}
C(r,t)=\frac{2}{R_0^2}\sum_{n=1}^{\infty}{\frac{J_0\left(\frac{\mu_n}{R_0}r\right)}{J_1\left(\mu_n\right)^2}
\e{-D \left(\frac{\mu_n}{R_0}\right)^2 t}\int_0^{R_0}{r' C_0(r')J_0\left(\frac{\mu_n}{R_0}r'\right)\text{d}r'}.
}
\label{eq:Solution}
\end{equation}

\noindent However, we can go one step further by substituting in our initial condition. In the case of out-diffusion,
\[
C_0(r)=\left\{
\begin{array}{lr}
	0 & r=R_0\\
	1 & \text{else}
\end{array}\right. .
\]

\noindent Upon noting that the value of a Riemann integral does not change upon altering a finite number of values of the integrand, we can set $C_0(r)=1$ everywhere. The integral can then be carried out and yields
\[
\int_0^{R_0}{r' J_0\left(\frac{\mu_n}{R_0}r'\right)\text{d}r'}=\frac{R_0^2 J_1\left(\mu_n\right)}{\mu_n},
\]

\noindent which makes our most specific solution
\begin{equation}
C(r,t)=2\sum_{n=1}^{\infty}{\frac{J_0\left(\frac{\mu_n}{R_0}r\right)}{\mu_n J_1\left(\mu_n\right)}
\e{-D \left(\frac{\mu_n}{R_0}\right)^2 t}.
}
\end{equation}

\noindent For in-diffusion one can either solve the inhomogeneous boundary problem ($C_\text{in}(R_0,t)=1$) or simply note that the diffusion equation remains invariant under $t'=-t$ when $D'=-D$, so that the solution only changes sign in the exponential, i.e. in- and out-diffusion are related by $C_\text{out}(r,t)+C_\text{in}(r,t) = 1$. We can now use this sufficiently specialized solution to make predictions for optimal loading conditions.

\section*{Appendix II: Notes on fiber production procedure}
These notes are meant to provide a guide to producing UV patch cords as pursued in this publication. Exact procedures strongly depend on the equipment in use; therefore, equipment specific values are not quoted.  The NIST recipe~\cite{Recipe} also has many useful hints, but the current version has a great deal of detail for steps we do not take and equipment we do not have.  Fiber connectorization is performed commercially.

\subsection*{Fiber preparation}
\emph{Pre-curing only}: After a small length of fiber has been cut from the reel (using e.g. scissors) and the cleaving tool is cleaned (as per manual), the fiber is stripped using a Thorlabs fiber buffer stripping tool and a razor blade. Depending on the fiber launching system used in later steps, one must take care not to remove too much of the coating. We use a Thorlabs MAX350D/M fiber launcher which will only clamp around the coated region. The exposed fiber is then cleaned using a lint-free cloth with a few drops of isopropyl alcohol. 

We employ a Fujikura CT-100 cleaver and Fujikura FH-100-400 fiber holders, which also provide a convenient means of protecting the bare fiber tips during transport and handling. Additionally, for high positional (and hence coupling) stability the fiber should be supported close to the facet. One side of the fiber is cleaved with an appreciable angle to avoid internal etalon effects. Some iteration on the cleaver settings to achieve a good angled cleave while maintaining output mode quality will be required. 

After a cleave the facet must be flat and without chips as observed by rotating the fiber under a microscope. Incorrect cleaver settings will rip the fiber rather than cutting which can cause helically shaped interfaces or ledges. Damage may also be caused to the hole structure, resulting in a brighter section of the fiber relative to areas where the holes are intact.   If the facet is satisfactory it can be secured e.g. in a spare FH-100-400 with the tip safely in the groove outside the magnetic holder.  

We use unloaded test pieces for pre-alignment and cleaver testing, which do solarize and are discarded afterward. Once this procedure can be completed within a few minutes, the desired length of fiber can be cut, flat cleaved on both sides and secured for transport to the loading facility. If a fusion splicer is available to collapse fiber ends after loading, sufficient coating has to be removed for reliable operation of the splicer.

\subsection*{Fiber Loading}
\emph{Pre-curing}: The cleaved and secured fiber is placed inside the pressure vessel, taking care in handling to prevent shattering of the tips or hole structure. The pressure lines and the vessel are purged with a dry gas (e.g. \Nt) to prevent water from migrating into the fiber if cold storage is anticipated after loading. It is not necessary to remove residual purging gas or air since the partial pressure will be very small and the diffusion coefficients substantially smaller than those of \Ht. The vessel is filled with \Ht\ at bottle pressure, $\sim\unit[130-180]{bar}$ at room temperature. Heating is not necessary due to the relatively fast in-diffusion speeds through the holes. However, if desired, the vessel can be placed in a suitable water-immersion heater. Once the requisite loading time is reached the vessel can be depressurized and the fiber removed.

Upon removal fibers may move to curing or may be placed in cold storage.  Care must be taken with cold fibers that condensation does not migrate into the hole structure to prevent ice-crystal formation and subsequent damage to the hole structure. We have thus employed sealed plastic containers and moisture absorbers. 
\\\\
\emph{Post-curing}: Loading patch cords proceeds similarly, with due consideration of the requisite minimum bending radii in the pressure vessel, particularly around the stiff connector/boot region.  Due to the reduced diffusion rate through the endcaps we heat the vessel to a maximum of $\unit[60]{^\circ C}$ to prevent damage to the polymers and glues employed in patch cord construction. The vessel should be cooled down before reconnecting to the \Ht\ manifold to prevent material mismatch from differential thermal expansion. If the fiber is stored cold after removal from the \Ht\ environment, the polymer parts can become brittle and need to be handled carefully in order to avoid breaking them.

\subsection*{Fiber Curing}
\emph{Pre-curing}: The fiber launcher setup is pre-aligned using test-segments of bare uncured fiber. Due to the relative challenge of finding the fundamental mode we employ a cage-mounting system to enhance rigidity between the coupling mirrors and the launcher optics. Detailed advice on bare fiber coupling in the context of UV curing can again be found in the appropriate section of \cite{Recipe}.

Prior to curing fiber facets are inspected and recleaved if required in order to improve mode-matching. Initial fiber alignment uses $\unit[5-10]{mW}$ of UV on the input of the fiber; if there is significant difficulty in coupling the fiber may be recleaved.  Once coupling is satisfactory, optical power is increased to the relevant curing levels. We typically cured open-ended fibers with relatively high powers of $\unit[100-200]{mW}$ on the input and coupling efficiencies of $\unit[50-60]{\%}$ such that curing proceeds fast relative to out-diffusion speeds.

Upon exposing the fiber to high power the output mode may shift and morph for a while. We have seen this on occasion with the output mode settling to the usual fundamental after several minutes. Exposure to high power can also cause the incoupling efficiency to drift over the course of tens of minutes either due to changes in the refractive index from curing or through heat dissipation (again acting on the refractive index and/or expansion of the fiber). Slight recoupling is sometimes required while the fiber settles.  

It has been pointed out that fibers with a flat cleaved collapsed region or endcap can rapidly form a Bragg grating in the endcap due to etalon effects which manifests in a rapid drop in transmission, some reflected light and high sensitivity of the fiber to gentle bending or slight temperature changes\cite{ColombeCommunication}. 
\\\\
\emph{Post-curing}: 
Fiber patch cords are stretched for approximately one day using a jig as described and straightened for curing.  The fiber is tensioned in order to force the jacketing to relax but too much tension will damage the connector region; a standing-wave acoustic vibration should be supported.  Since the fiber will likely still not be fully relaxed at the time of curing we have designed and built a stretching jig consisting of a $\unit[25]{cm}$ central cylinder and two equal diameter $\unit[90]{^\circ}$ deflector sections. All three parts are movable over a short distance such that the total length of the movable part equals half a revolution of the fiber around the central cylinder. This allows for a fiber of any length (limited by available space on the optics table that is) to be stretched. 

We employ purged collimators in a pre-aligned optical setup for UV curing.  For coupling and transmission the connector region must be straight; this is determined iteratively by observing the fluorescence of the jacketing and fiber coating due to UV illumination.  The fiber is tensioned and twisted while fiber transmission is monitored. Once a stable configuration has been found, curing can proceed as above.  Additional precaution must be taken to ensure good coupling efficiency as the fiber ferrule facet may be irreversibly damaged by high power UV.  An FC-connector is usually quoted with being able to dissipate $\unit[500]{mW}$ of power, but UV absorption in the used materials is not accounted for in these statements.

Since out-diffusion of \Ht\ through the endcaps at room temperature is a much smaller concern than with open-ended fibers, we cure at lower powers for convenience ($\unit[20-30]{mW}$ of UV on the output over a day). After curing has finished fibers are stretched again for days to weeks until transmission appears independent of fiber conformation/coiling.

\section*{Funding}
ARC Centre of Excellence for Engineered Quantum Systems (EQuS) (CE110001013), US Intelligence Advanced Research Projects Activity through the Army Research Office (ARO) (W911NF-16-1-0070), and a private grant from H. and A. Harley.

\section*{Acknowledgments}
The authors wish to thank Kevin Cook and Prof. John Canning (iPL) for the use of their hydrogen vessel and loading facilities and Dr. Cook for his great help and input with \Ht\ loading. We would also like to thank Cato Fagermo from \nkt\ for technical advice, Andrew C. Wilson for his input with the $\unit[313]{nm}$ laser system and both him and in particular Yves Colombe for their advice and useful discussions about their work on bare fibers.

\end{document}